# Three-Dimensional Stateful Material Implication Logic

Gina C. Adam[1], Brian D. Hoskins[2], Mirko Prezioso[1], and Dmitri B. Strukov[1]

## Abstract

Monolithic three-dimensional integration of memory and logic circuits could dramatically improve performance and energy efficiency of computing systems.[1,2] Some conventional and emerging memories are suitable for vertical integration,[3,4] including highly scalable metal-oxide resistive switching devices ("memristors"),[5-8] yet integration of logic circuits proves to be much more challenging.[1,9,10] Here we demonstrate memory and logic functionality in a monolithic three-dimensional circuit by adapting recently proposed memristor-based stateful material implication logic.[11] Though such logic has been already implemented with a variety of memory devices,[12-16] prohibitively large device variability in the most prospective memristor-based circuits has limited experimental demonstrations to simple gates and just a few cycles of operations. By developing a low-temperature, low-variability fabrication process, and modifying the original circuit to increase its robustness to device imperfections, we experimentally show, for the first time, reliable multi-cycle multi-gate material implication logic operation within a three-dimensional stack of monolithically integrated memristors. The direct data manipulation in three dimensions enables extremely compact and high-throughput logic-in-memory computing[17-20] and, remarkably, presents a viable solution for the Feynman's grand challenge of implementing an 8-bit adder at the nanoscale[21].

---

[1]Department of Electrical and Computer Engineering, University of California at Santa Barbara, Santa Barbara, CA 93106. [2]Department of Material Science, University of California at Santa Barbara, Santa Barbara, CA 93106. Correspondence and requests for materials should be addressed to G. C. A. and D. B. S. (email: gina_adam@engineering.ucsb.edu, strukov@ece.ucsb.edu ).

Material implication (IMP) is a universal Boolean logic (Fig. 1a) particularly suitable for implementing "stateful" logic circuits.[11] At the core of stateful logic are memory devices which serve a dual role - performing computation and storing (latching) the results. The most prospective implementation is based on highly-scalable memristors.[22-24] In the simplest case, memristors are two-terminal devices, whose conductance can be switched reversibly with relatively large (write) voltages, e.g. applying $V \gtrsim V_{set}$ to switch device to the ON state characterized by high conductance $G_{ON}$, and $V \lesssim V_{reset}$ to switch it to the OFF state with low conductance $G_{OFF}$ (Fig. 1b). The device's conductance remains unchanged when relatively small (read) voltages are applied. Specifically, in one realization of memristor-based IMP logic, logic states '0' and '1' are encoded with low and high conductive states of a memristor. Using a divider circuits shown in Fig. 1c, $q' \leftarrow p$ IMP $q$, that is an implication between logic variables $q$ and $p$, stored in memristors Q and P, respectively, is performed by applying specific "clock" voltage pulses $V_P$ and $V_L$, so that the result of the computation is placed in Q as a new conductive state. Similar to other nonconventional computing approaches,[17,18] voltage pulses $V_P$ and $V_L$ are effectively clock signals which do not carry any information. Their amplitudes are fixed and are chosen according to the load conductance $G_L$, and memristor parameters, e.g. $G_{ON}$, $G_{OFF}$, $V_{set}$, and $V_{reset}$ for the ideal memristor without variations (Fig. 1b), such that the device Q switches from high to low conductive state only when device P is in the low conductive state.

The appealing feature of stateful logic is that the result of the logic operation is immediately latched. Thus, IMP logic circuits based on non-volatile memristors are immune to shortages in power supply, which could be advantageous in the context of energy scavenging applications. Even more importantly, stateful logic does not draw static power and enables very high throughput information processing due to the possibility of fine-grained pipelining. In many respects, stateful IMP logic is similar to other logic-in-memory computing approaches,[17-20] which does not suffer from the memory bottleneck problem of conventional Von Neumann architectures.[25] Several theoretical studies have predicted significantly higher performance and energy-efficiency for memristor-based IMP logic circuits and very similar concepts over conventional approaches for high-throughput computing applications.[26,27]

However, even simple experimental demonstrations of memristor-based IMP logic are challenging due to a memristor's cycle-to-cycle and device-to-device variations. Device variations reduce allowed range of $V_P$ and $V_L$ voltages within which correct operation is assured. In fact, IMP logic is more prone to variations and demonstration of memory functionality does not guarantee that the same circuit can be adapted for performing logic operations (Sec. 3 of Supplementary Information). Extending IMP logic to three dimensional circuits is even more difficult, demanding more sophisticated fabrication processes and higher integration density which can aggravate the device variation problems. The main goal of this paper is to address these challenges and ultimately demonstrate robust stateful IMP logic in monolithic three-dimensional metal-oxide memristor structures.

The fabricated circuit consists of a two-level stack with four metal-oxide memristors. Two memristors were fabricated in the bottom level, and two others were monolithically integrated directly above, with all devices sharing a common middle electrode (Fig. 2a-c). The major steps involved in fabrication were: patterning of Ta/Pt bottom electrode by e-beam evaporation and lift-off; patterning of bottom device's $Al_2O_3/TiO_{2-x}$ layer and Ti/Pt middle electrode by reactive sputtering and lift-off; planarization by chemical mechanical polishing and etch-back of plasma-deposited sacrificial silicon oxide; and, patterning of top device's $Al_2O_3/TiO_{2-x}$ layer and Ti/Pt top electrode by reactive sputtering and lift-off (Fig. 2d-g). The device structure, oxide film thicknesses, and titanium oxide stoichiometry, which was controlled by changing oxygen to nitrogen flow ratio during sputtering, were selected based on our earlier study,[28] with the primary objective of lowering forming voltages and improving uniformity of switching characteristics.

In particular, thin Ti and Ta layers were deposited to improve electrode adhesion. Addition of Ti to the middle and top electrodes also ensured ohmic interfaces with the titanium dioxide layer, which was important for the device's asymmetry.[29] Low forming voltages reduced electrical stress during electroforming,[28] while in-situ contacts between titanium oxide and the metal electrodes, which were fabricated without breaking the vacuum, ensured high-quality interfaces,[30] with both factors essential for improving uniformity of memristor's switching characteristics. Furthermore, planarization reduced middle electrode roughness that resulted from residual sidewall deposition and was critical for lowering variations in top-level devices (Fig. S1-S3). The absence of annealing step, which is typically used for fine-tuning of the defect profile in metal oxide memristors, [24,28] and the low-temperature fabrication budget with temperatures below 300ºC during the sputter deposition, simplified three-dimensional integration and makes the fabrication process compatible with conventional semiconductor technology. More details on fabrication are provided in Section 1 of Supplementary Information.

Figure 3a,b shows typical memristor $I$-$V$ characteristics obtained by applying positive and negative quasi-DC triangular voltage sweeps. Switching polarities for all devices correspond to the bottom active interface, which is in agreement with the devices' asymmetric structure. For all devices, the set switching is rather sharp, while the reset process is gradual. For example, for the device B1 reset transition starts at $V_{reset}^{min} \approx$ -1.5 V; however, to avoid partial switching, voltage exceeding $V_{reset}^{max} \approx$ -2.2 V must be applied (Fig. 3b). A slightly thicker titanium dioxide layer for the bottom devices resulted in higher set threshold voltages as compared to those of the top ones (Figs. 3a and S4). As Figures 3c and d show, repetitive switching between ON and OFF states of one device did not disturb the state of others, thus suggesting that thermal crosstalk is negligible. Ratio of currents measured at 0.1 V between the ON and OFF states were well above one order of magnitude for all memristors. Other characteristics, such as endurance and retention, were close to those reported earlier for similar devices.[28]

Significant set threshold voltage variations (Fig. 3b) is a major challenge for implementing IMP logic. Therefore, it is natural to choose circuit parameters (i.e. $G_L$, $V_L$, $V_P$) that maximize the range of variations, also referred as margins, which can be tolerated without comprising the correctness of logic operation. Some earlier works suggested choosing $G_L$ = $(G_{ON}G_{OFF})^{1/2}$ for the most optimal design,[31] however, our simple analysis of IMP logic operation (Sec. 3 of Supplementary Information) shows that set margins monotonically increase as the load conductance decreases (Fig. 1e, f). The largest margins are for $G_L$ = 0, which cannot be implemented with the original circuit, though can be easily realized by replacing the load resistance and voltage source with a current source (Fig. 1d). The transition from the original circuit with earlier suggested $G_L$ to the modified one with an optimized current source $I_L$ increased set margins by more than 20% (Fig. 1e). Such a boost in variation tolerance was critical for our experiment by allowing it to cope with virtually all experimentally observed variations (Fig. S5). It should be noted that, in principle, IMP logic can also be implemented using a memristor's reset transition, i.e. assuming that logic states "0" and "1" are represented by the ON and OFF states instead. However, this would not be helpful in our case, because of the gradual reset transition– see Section 3 of Supplementary Information for more details.

Using the variation tolerant design with optimal values of $I_L$ and $V_P$, which were obtained from accurate numerical simulations based on experimental (nonlinear) $I$-$V$ curves, we successfully demonstrated IMP logic with the fabricated memristor circuit (Figs. 4 and 5). In the first set of experiments, a series of IMP operations were performed sequentially utilizing four different pairs of memristors (Figs. 4 and S7). Before each logic operation, the devices were always written to the specified initial states, therefore this experiment is a proof of memory and logic functionality implemented within the same circuit. Moreover, the considered pairs constitute all possible combinations of memristor's polarities in IMP circuit and hence are sufficient to compute and move information in any direction within the circuit.

In most cases of the first experiment, output conductances are close to the extreme ON and OFF values, so that it should be possible to cascade IMP logic gates, i.e. use the output of one gate as an input for the other. To confirm this, in the next series of experiments, we implemented NAND Boolean logic operation, for which inputs were the states of the bottom level devices and the output was stored in one of the top level memristors (Fig. 5). The NAND gate was realized in three steps - an unconditional reset, followed by two sequential IMP operations.[11] The result of the first IMP operation was stored in the top level device, which was then used as one of the inputs to the second IMP gate. In some rare cases ($\sim$ 6.5% of total IMP operations), there is some visible reduction in the ON-to-OFF conductance ratio. This is not desirable because set margins decrease with ON-to-OFF ratio (Fig. 1e). One plausible solution to restore the ratio is to read the state and write it back, i.e. similar to what was implemented in the first experiment.

Interestingly, three-dimensional IMP logic enables a practical solution for one of the Feynman Grand Challenges – the implementation of an 8-bit adder which fits in a cube no larger than 50 nanometres in any dimension.[21] The major building block – a full adder, which adds Boolean variables $a$, $b$, and $c_{in}$ to calculate sum $s$, and carry-out $c_{out}$, requires 6 memristors and consists of two monolithically stacked 2×2 crossbars sharing the middle electrodes (Fig. 6a). Two of the memristors in the crossbar are assumed to be either not formed or always kept in the OFF state (Fig. 6b), which eliminates leakage currents typical for crossbar circuits and makes IMP logic set margins similar to those of the demonstrated circuit. In particular, at the start of computation, $a$, $b$, and $c_{in}$ are written to the specific locations in the circuit (Fig. 6c). A sequence of NAND operations, each consisting of one unconditional reset step and two IMPs (Fig. 5), is then performed to compute $c_{out}$ and $s$ according to the particular implementation of

Fig. 6d. An occasional NOT operation is implemented with one unconditional reset step and one IMP step and is used to move variables within the circuit. In total, the full adder is implemented with 9 NAND gates and 4 NOT gates, i.e. 13 unconditional reset steps and 22 IMP steps. The simplest way to read an output of an adder is to measure electrically the state of memristors T2 and T3 (Fig. 6c). Alternatively, the output can be sensed as a mechanical deformation of upper metal electrodes, which is often observed in metal-oxide memristors,[32] or using scanning Joule expansion microscopy.[33] Finally, a full 8-bit adder could be implemented in a ripple-carry style[34] by performing full adder operation 8 times.

In summary, we have demonstrated logic-in-memory computing in three-dimensional monolithically integrated circuits. As the memristor technology continues its rapid progress and will eventually become sufficiently advanced to enable large-scale integration of memristive devices with sub-nanosecond, pico-Joule switching with $>10^{14}$ cycles of endurance, which so far was demonstrated for discrete devices,[22-24] we expect that the presented approach will become attractive for high-throughput and memory-bound computing applications suffering from memory bottleneck problems. Furthermore, we showed how the presented approach establishes a realistic pathway towards resolving one of the Feynman's Grand Challenges. The remaining challenge is to scale down the circuitry (Fig. 6a), which does not seem unrealistic task given that discrete metal-oxide memristors with similar dimensions[24] and much more complex (but less dense) memristive circuits have been already demonstrated. [4,5,7,28]

**Acknowledgments**

We acknowledge useful discussions with F. Alibart, F. Merrikh-Bayat, B. Mitchell, J. Rode, and B. Thibeault. This work was supported by the AFOSR under the MURI grant FA9550-12-1-0038, by DARPA under Contract No. HR0011-13-C-0051UPSIDE via BAE Systems, and by the Department of State under the International Fulbright Science and Technology Award.

**Author Contributions**

G.C.A., B.D.H., and D.B.S. designed research. G.C.A. and B.D.H. performed fabrication, device characterization, and circuit experiments. M.P. advised on the device fabrication. D.B.S. advised on all parts of the project. All discussed and interpreted results. G.C.A. and D.B.S. wrote the manuscript.

**Figure captions**

**Figure 1.** Memristor-based material implication logic: (a) Logic truth table and its mapping to memristor's states. (b) A sketch of simplified (linear) $I$-$V$ switching curve for a memristor. The thick (thin) solid lines show schematically an $I$-$V$ curve with average (maximum and minimum) set and reset thresholds. The inset shows experimental setup. (c) Originally proposed[11] and (d) modified IMP logic circuits with particular polarity of memristors. (Other possible configurations are shown in Fig. S6.) (e) The set margins as a function of load conductance for several representative ON-to-OFF conductance ratios. For convenience, margins and load conductances are normalized with respect to mid-range set voltages $V^*_{set}$ and $G_{ON}$, respectively. Solid dots show margins for previously proposed optimal load conductance $G_L$', while solid triangles are margins which were obtained with numerical simulations using experimental device characteristics. The solid and dashed horizontal lines denote the maximum and the actual set margins, respectively, when taking into account experimental data. (f) A diagram showing definition of margins in the context of set transition.

**Figure 2.** The $Al_2O_3/TiO_{2-x}$ memristor circuit: fabrication details. (a) An equivalent circuit. B1 and B2 denote bottom devices, while T1 and T2 the top ones. (b) A cartoon of device's cross-section showing the material layers and their corresponding thicknesses. (c) A top-view scanning-electron-microscope image of the circuit. The red, blue, and purple colours were added to highlight the location of bottom and top devices, and their overlap, respectively. (d-e) A top-view atomic-force-microscope images of the circuit during different stages of fabrication, in particular showing: (d) bottom electrode; (e) middle electrode; (f) middle electrode after planarization step; and (g) top electrode.

**Figure 3.** The $Al_2O_3/TiO_{2-x}$ memristor circuit: electrical characterization. (a) Representative $I$-$V$ curves for all devices. (b) Switching $I$-$V$s showing 100 cycles of operation for the device B2 and the corresponding cycle-to-cycle set switching voltage statistics. (c) Conductance of the device B1 that was repeatedly switched 200 times and (d) those of the other three devices in the circuit that were kept in the OFF states for the first 100 cycles, and then in the ON states for the remaining 100 cycles. In all experiments, the memristors were switched by applying triangular voltage pulses to the corresponding top terminal of the device.

**Figure 4.** Material implication logic results. (a-d) Circuit schematics, and (e-i) corresponding experimental results showing device's conductances before and after IMP operation

implemented with various initial states and pairs of memristors in a circuit. On panels (e-i) each graph shows the averaged conductances and their standard deviations for 20 experiments. IMP logic was performed by biasing device $V_P = 0.25$ V and applying a 10-ms $I_L = 550$ μA load current pulse for the cases on panels (a, d), i.e. when the result was written into the bottom device, and $I_L = 200$ μA when the output was one of the top devices (panels b, c).

**Figure 5.** NAND Boolean operation via material implication logic. (a) Schematics and truth table showing intermediate steps. (b) Experimental results showing 80 cycles of operation with >93% yield for all four combinations of initial states. The initial states were set similarly to the Figure 4 experiments, while $V_P = -0.15$ V, and load current was applied as 10-ms pulse with $I_L = -550$ μA.

**Figure 6.** A full adder implementation with 3D IMP logic: (a) Cartoon of a structure and (b) its equivalent circuit. (c, d) A sequence of steps and specific mapping of logic variables to the circuit's memristors for a particular implementations of full adder shown on panel d. The last step on panel d, in which $c_{out}$ is placed in the same location as $c_{in}$, is only required to ensure modular design, but might be omitted in more optimal implementations.

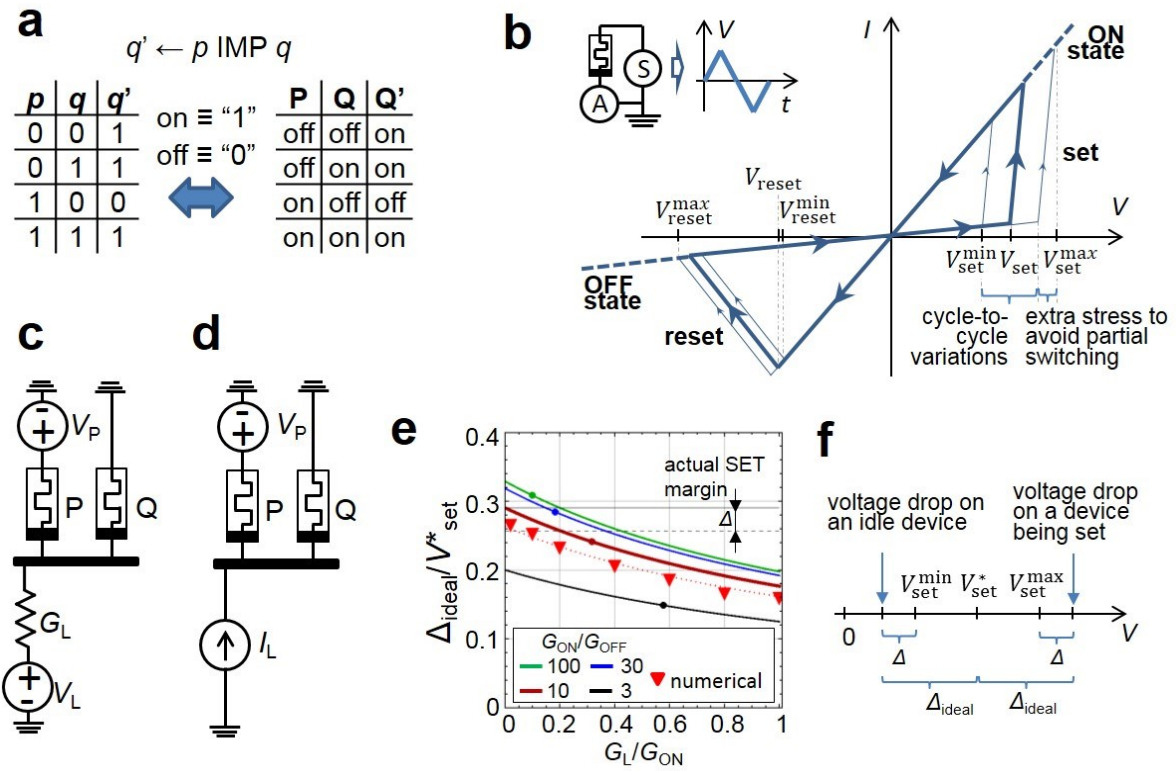

Figure 1

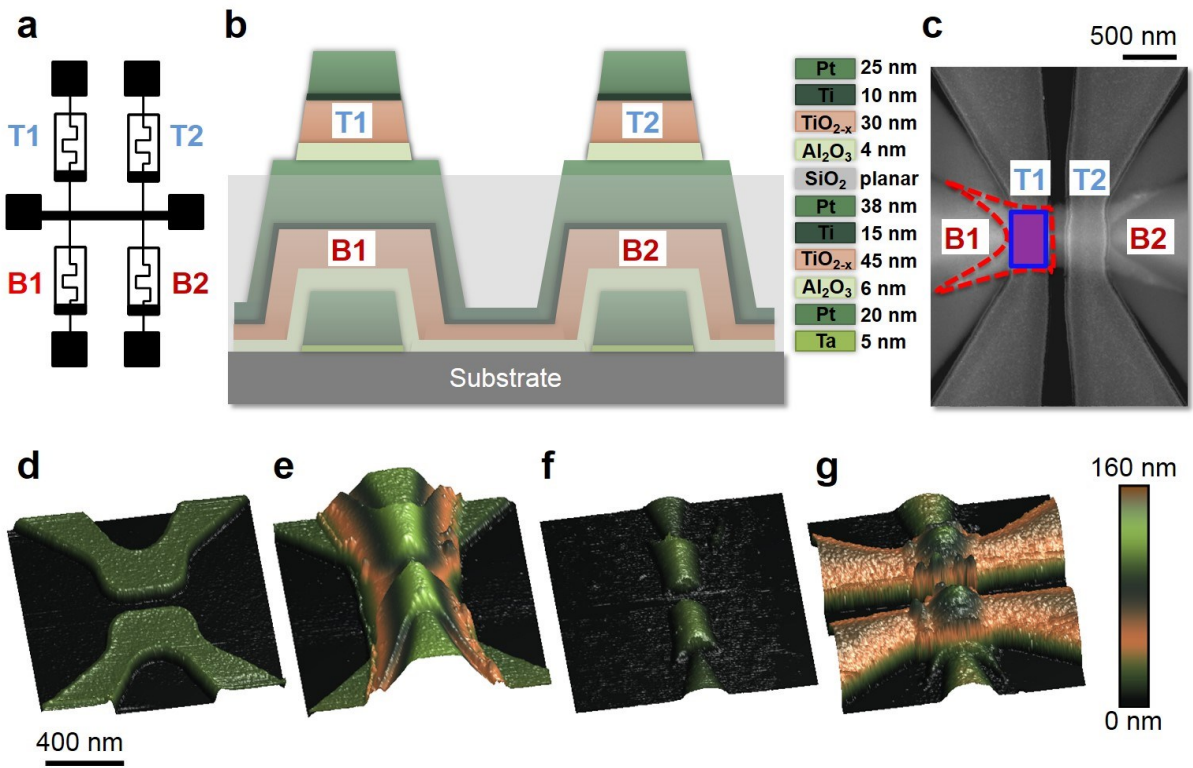

Figure 2

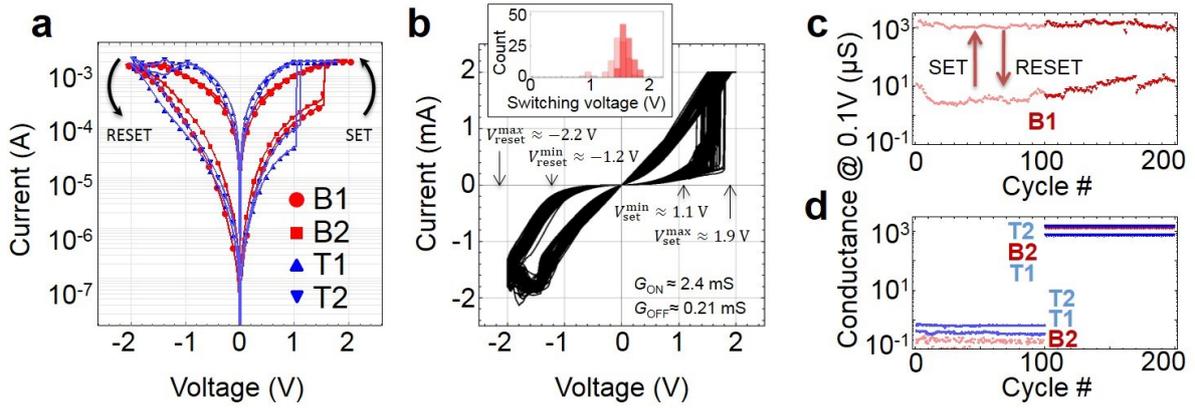

Figure 3

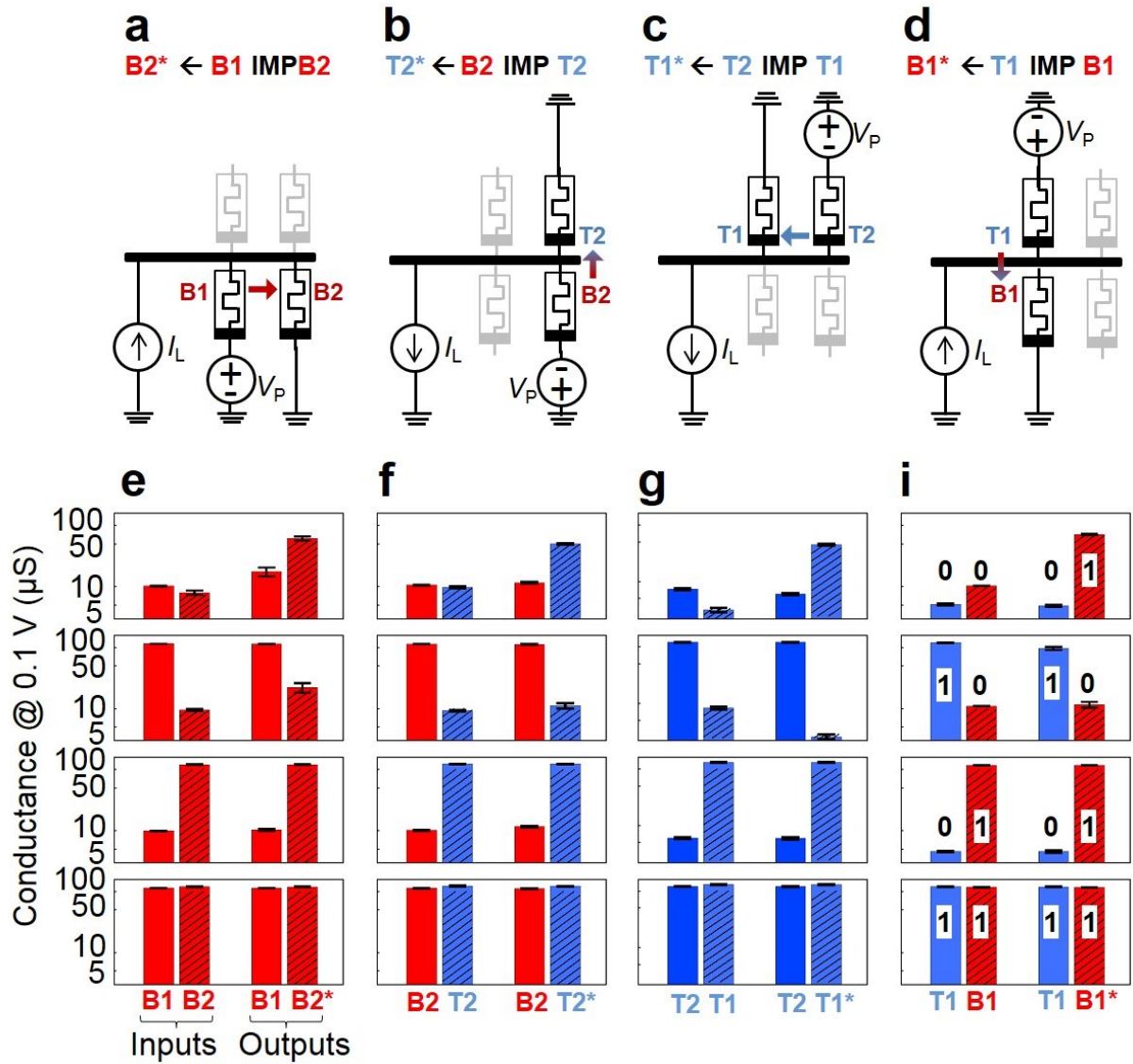

Figure 4

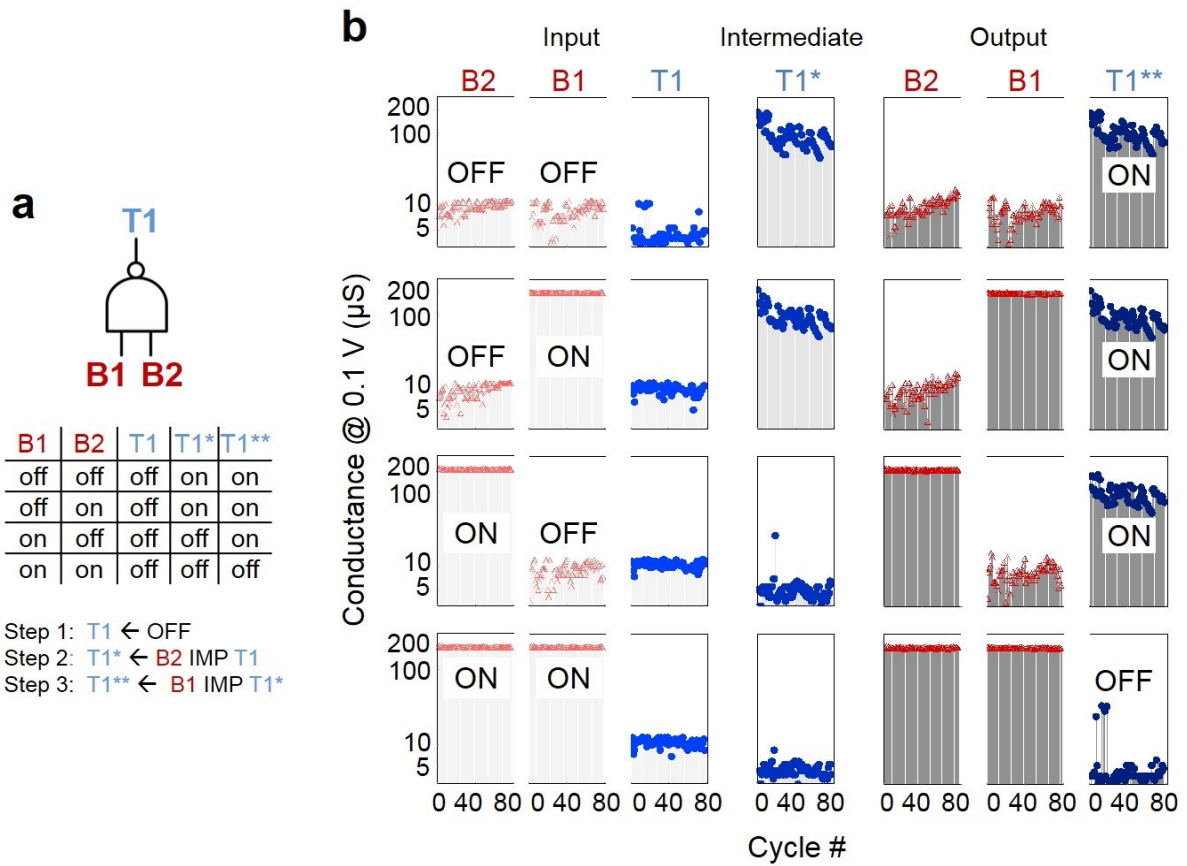

Figure 5

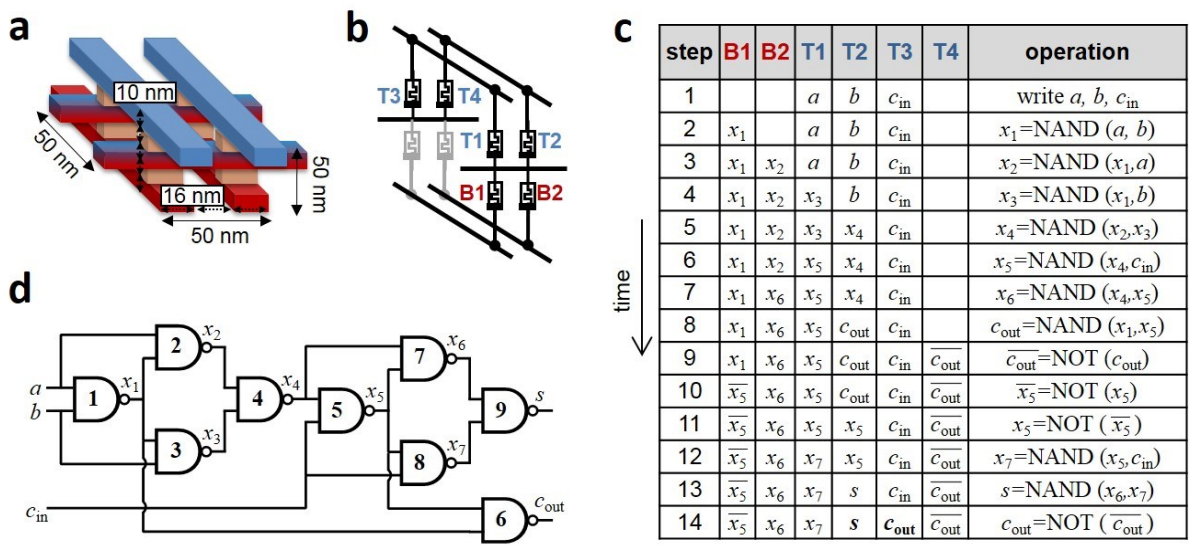

Figure 6

# Supplementary information

## 1. Circuit fabrication

Devices were fabricated on a Si wafer coated with 200 nm thermal $SiO_2$. Circuit fabrication involved four lithography steps using an ASML S500 / 300 DUV stepper with a 248 nm laser. To prevent from misalignment of device layers, the bottom devices were made larger with an active area of 500 nm × 500 nm, as compared to a 300 nm × 500 nm active area of top devices.

In particular, in the first lithography step the bottom electrode was patterned using a developable antireflective coating (DSK-101-307 from Brewer Science, spin speed 2500 rpm, bake 185ºC, thickness ~50 nm) and positive photoresist (UV210-0.3 from Dow, spin speed 2500 rpm, bake 135ºC, thickness ~300 nm). 5 nm / 20 nm of Ta / Pt were evaporated at 0.7 A/sec deposition rate in a thin film metal e-beam evaporator. After the liftoff, a "descum" by active oxygen dry etching at 200ºC for 5 minutes was performed to remove photoresist traces.

In the next lithography step, the middle electrode was patterned and the bottom device layer (6 nm / 45 nm of $Al_2O_3$ / $TiO_{2-x}$ bi-layer) and middle electrode (15 nm / 38 nm of Ti / Pt) were deposited using low temperature (< 300ºC) reactive sputtering in an AJA ATC 2200-V sputter system. To minimize sidewall redeposition on the photoresist, which was undercut during sputtering of the middle electrode and caused "bunny-ear" formation around the edges of middle electrode (Fig. S1a), both metals were deposited at 0.9 mTorr, which is the minimum pressure needed to maintain plasma in the sputtering chamber. Also, the thickness of the photoresist undercut layer was optimized to provide more shadowing by using a liftoff layer of LOL2000 (from Shipley Microposit, spin speed 3500 rpm, bake 210ºC, thickness ~200 nm) followed by the same DSK101/ UV210 stack as for the first lithography step mentioned above. Occasional lumps were reduced to the height of ~ 20-30 nm by swabbing in isopropanol (Fig. S1b).

Severe topography of the bottom level devices (Fig. 2e) may cause shorts and large variations in top level devices. To overcome this problem, a planarization step was performed using chemical mechanical polishing and etch-back of 750 nm of sacrificial $SiO_2$. $SiO_2$ served the double purpose: as a sacrificial material for planarization and as an insulation among devices. The most optimal planarization was achieved by depositing $SiO_2$ at 175ºC using PECVD. Following the deposition, 400 nm of SiO$_2$ were removed by chemical mechanical polishing for 3 min achieving surface roughness of less than 1 nm. The last step in the planarization procedure was to etch back ~ 250 nm of SiO$_2$ until the middle electrodes were exposed (Fig. 2f). Several etch-back approaches were investigated with the best results achieved using CHF$_3$ at 50 W, which had an etch rate of 0.2 nm/s (Fig. S2). In particular, the dry-etching with CHF$_3$ was done in steps to ensure < 5 nm roughness in the exposed middle electrode. AFM scans were performed after each etching step to check the thickness of the exposed electrode (Fig. S3) and to confirm that the post-etch surface has no traces of bunny-ear formations.

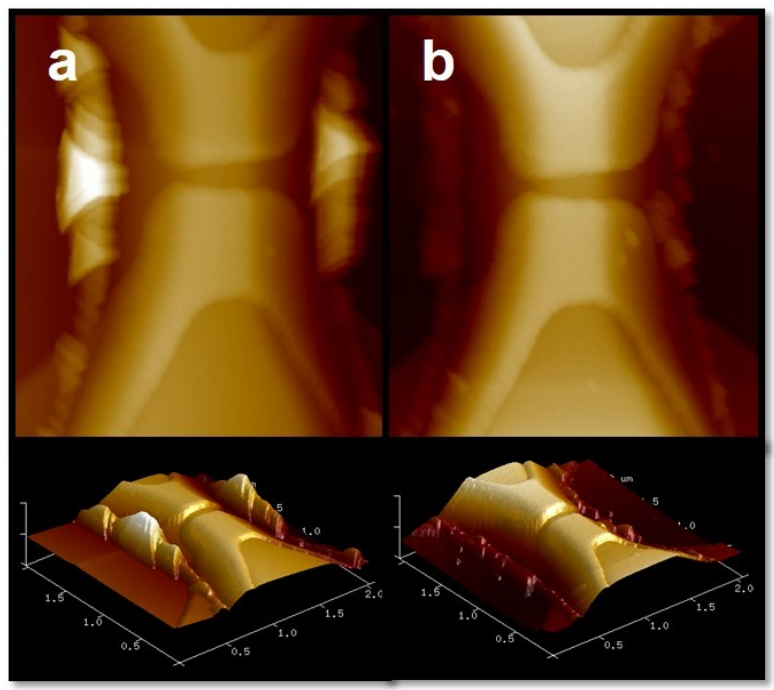

**Figure S1.** Middle electrode topology due to sidewall redeposition during sputtering (a) using standard process which results in > 200 nm lumps at the edges of the electrode and (b) after deposition optimization and swabbing method, which allows reduction of lumps to 20-30 nm.

After planarization and partial middle electrode exposure, the top layer devices were completed by in-situ reactive sputtering of the switching layer, which consisted of 4 nm / 30 nm of Al$_2$O$_3$ /TiO$_{2-x}$, and Ti (15 nm) / Pt (25 nm) top electrode over patterned photoresist (DSK101/UV210). No oxygen descum was performed before deposition in order to avoid potential oxidation of the bottom switching layer and to maintain TiO$_{2-x}$ stoichiometry.

Lastly, the pads of the bottom and middle electrodes were exposed through a CHF$_3$ etch of the sacrificial SiO$_2$ which was used for planarization.

In all lithography steps, the photoresist was stripped in the 1165 solvent (from Shipley Microposit) for 24 h at 80ºC.

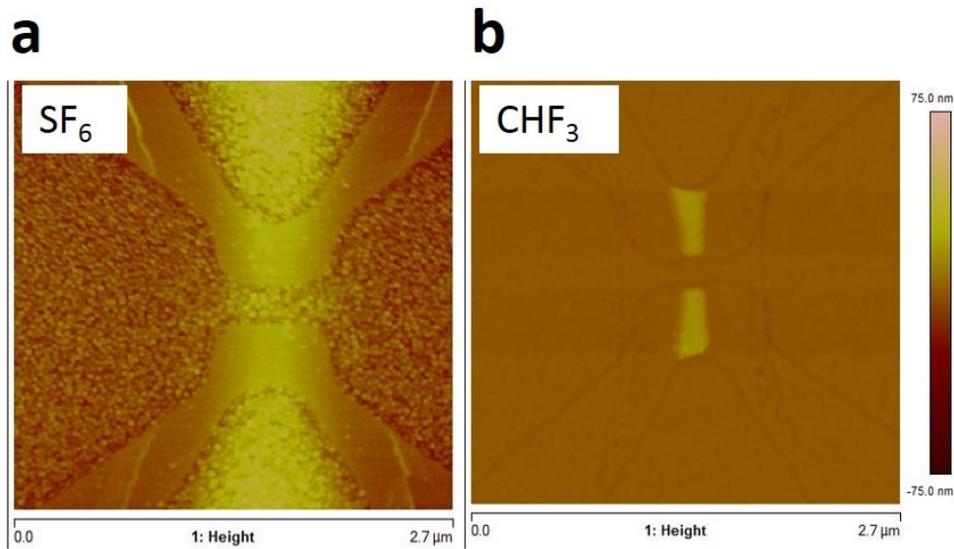

**Figure S2.** Comparison of two etch back methods for $SiO_2$. (a) SF6 achieving quadratic mean surface roughness > 6 nm and (b) $CHF_3$ with roughness less than 1 nm.

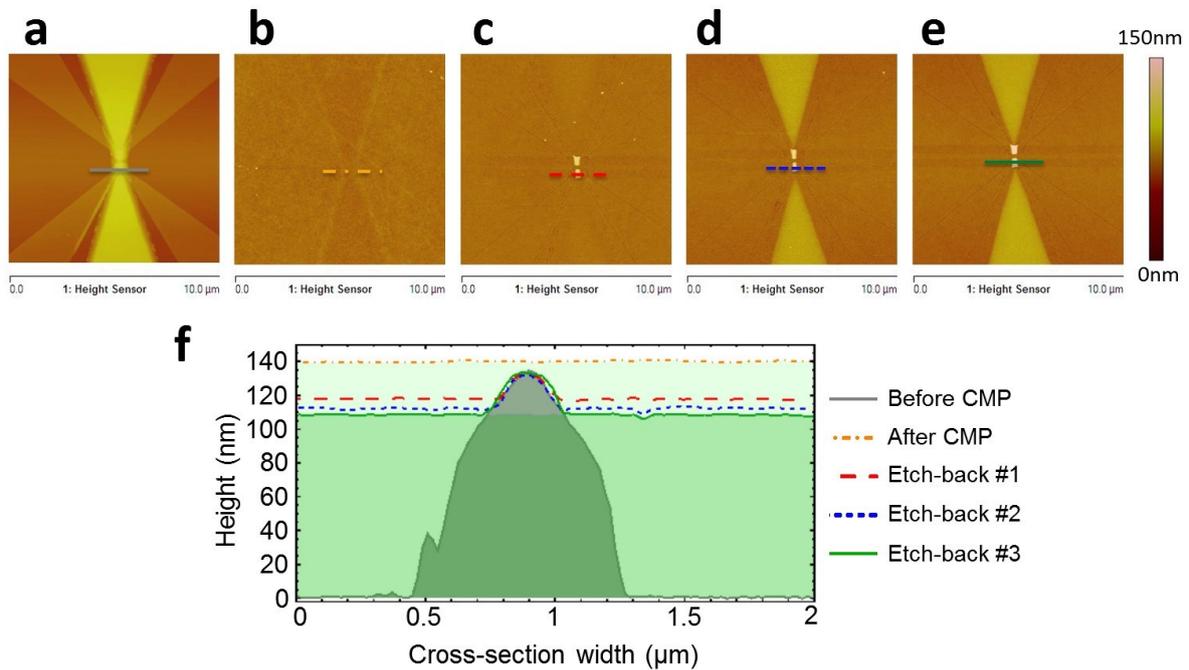

**Figure S3.** A top-view AFM images of the circuit during different stages of planarization, in particular showing: (a) bottom device before planarization; (b) after chemical-mechanical polishing of $SiO_2$ deposited over bottom device; (c) after etch #1 using $CHF_3$ for 1200 s showing partially exposed 18-nm-high middle electrode; (d) after etch #2 using $CHF_3$ for 20 s showing partially exposed 22-nm-high middle electrode; (e) after etch #3 using $CHF_3$ for 20 s showing partially exposed 28-nm-high middle electrode. (f) AFM height profiles taken across middle portion of the device (see marks on panels a-e) at the different planarization stages.

## 2. Electrical testing and device forming

All electrical testing was performed with an Agilent B1500 tool. The memristors were electroformed by grounding the device's bottom electrode and applying a current-controlled quasi-DC ramp-up to the device's top electrode, while keeping all other circuit terminals floating. For most of the devices forming voltages were around ~ 2-3 V, while device T1 did not require forming (Fig. S4). To minimize current leakage during the forming process, each memristor was switched to the OFF state immediately after forming.

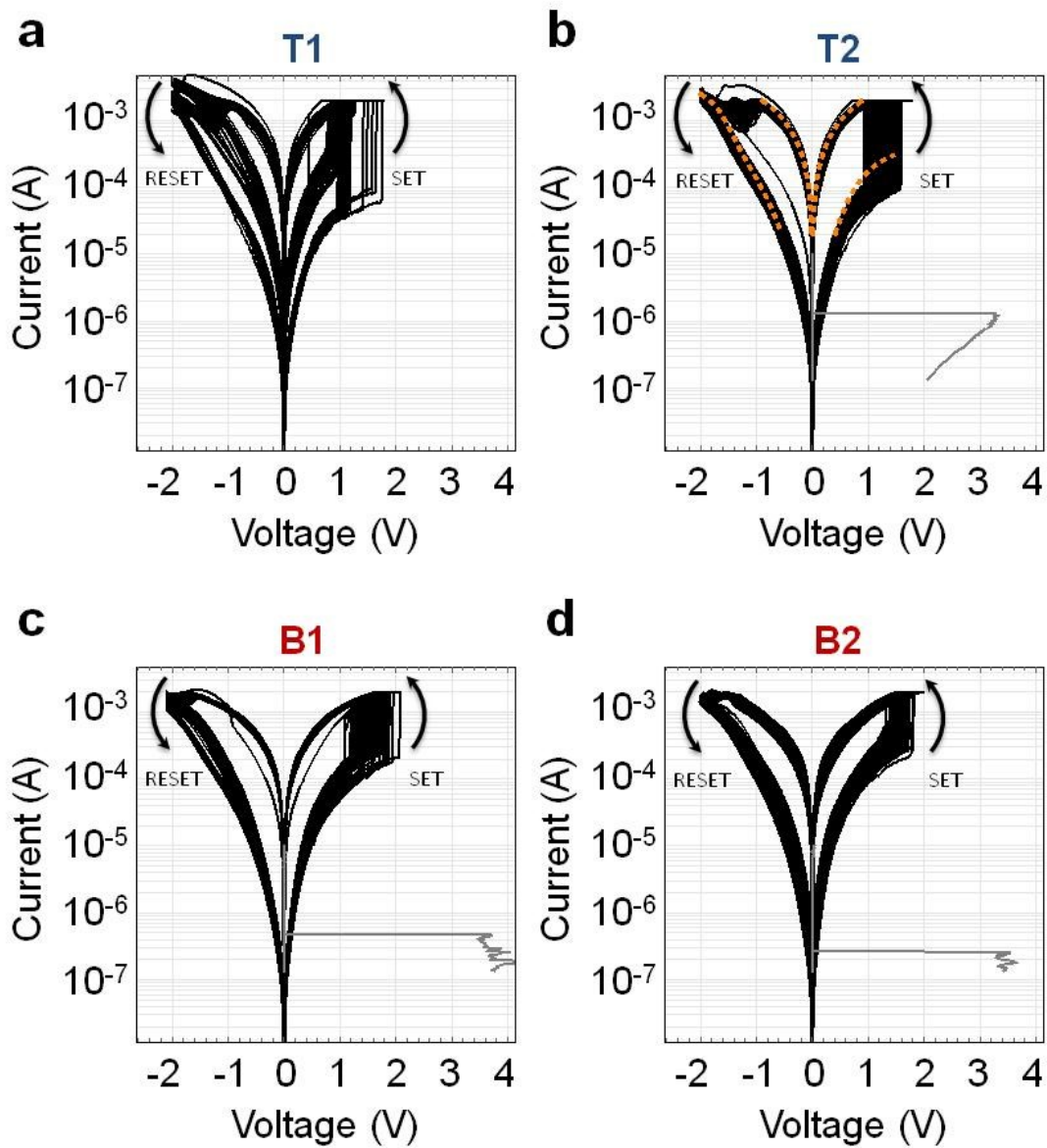

**Figure S4.** (a-d) *I-V* curves showing 100 cycles of switching for all devices. Gray lines show current-controlled forming *I-V*s. The dashed orange curve on panel b is a fitting used for the numerical simulations. For all cases, quasi-DC triangular voltage sweep was applied to the corresponding top terminals of the devices.

For all devices the most severe are cycle-to-cycle variations in set transition (Fig. S5), which range from 0.7 V to 1.6 V for the top layer devices, and from 1.1 V to 1.9 V for the bottom ones. However, because of gradual switching, $|V^{\max} - V^{\min}|$ statistics is comparable or wider for reset transition (Fig. S5).

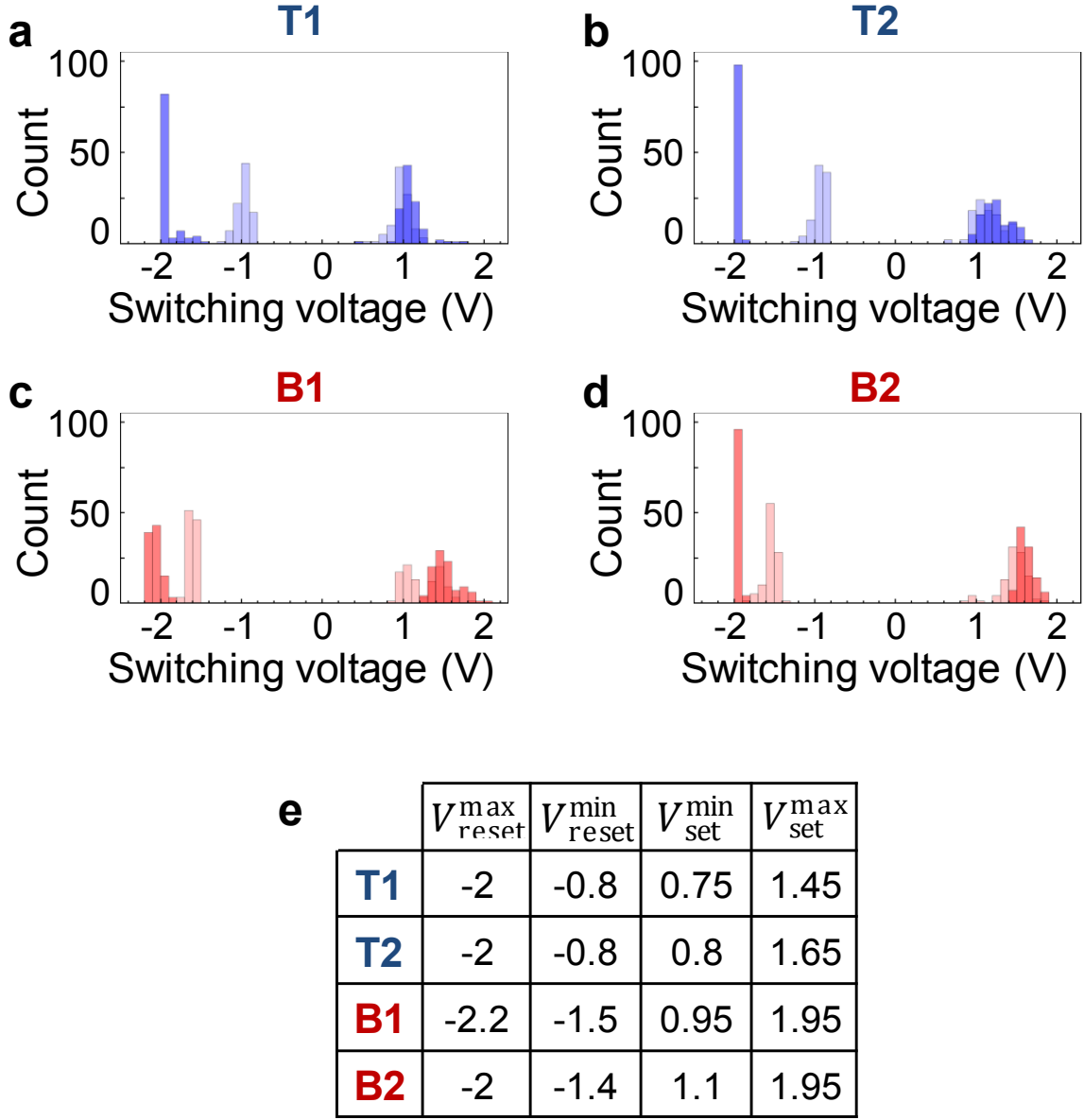

| e | $V_{\text{reset}}^{\max}$ | $V_{\text{reset}}^{\min}$ | $V_{\text{set}}^{\min}$ | $V_{\text{set}}^{\max}$ |
|---|---|---|---|---|
| **T1** | -2 | -0.8 | 0.75 | 1.45 |
| **T2** | -2 | -0.8 | 0.8 | 1.65 |
| **B1** | -2.2 | -1.5 | 0.95 | 1.95 |
| **B2** | -2 | -1.4 | 1.1 | 1.95 |

**Figure S5.** Switching voltages statistics extracted from experimental results shown on Fig. S4 for (a) T1, (b) T2, (c) B1, and (d) B2 devices. On all panels, light and dark colors show the $V^{\min}$ and $V^{\max}$ voltage distributions, correspondingly, for set and reset transitions. For the set transition, the switching occurs as sequence of abrupt changes in current and $V^{\min}$ ($V^{\max}$) is defined as the voltage of the first (last) abrupt change. The reset transition is more gradual and here $V^{\min}$ ($V^{\max}$) is calculated as the voltage at which the change in $I$-$V$ curvature is the largest (smallest) near the offset (end) of switching. (e) Table summarizing key parameters.

### 3. Material implication logic

The optimal circuit parameters $V_P$, $V_L$ and $G_L$, which result in the largest set margins could be derived analytically for the memristors with linear $I$-$V$ (Fig. 1b). Let us first consider an IMP circuit with a specific "parallel" configuration of memristors (Figs. 1c and S6a). Assuming for convenience that $V_Q = 0$, the proper operation of the material implication logic circuit shown on Figs. 1a, c require that device Q is set only when both P and Q are in the OFF state, i.e.

$$-V_C|_{P=OFF,Q=OFF} \geq V_{set}^{max} \qquad (1)$$

$$-V_C|_{OTHERS} < V_{set}^{min} \qquad (2)$$

where

$$V_C = \frac{-G_L V_L - G_P V_P}{G_L + G_P + G_Q} \qquad (3)$$

is a voltage on the common electrode. Device P should not be disturbed during the IMP operation, i.e.

$$(V_P - V_C)|_{ANY} < V_{set}^{min} \qquad (4)$$

$$(V_P - V_C)|_{ANY} > V_{reset}^{min} \qquad (5)$$

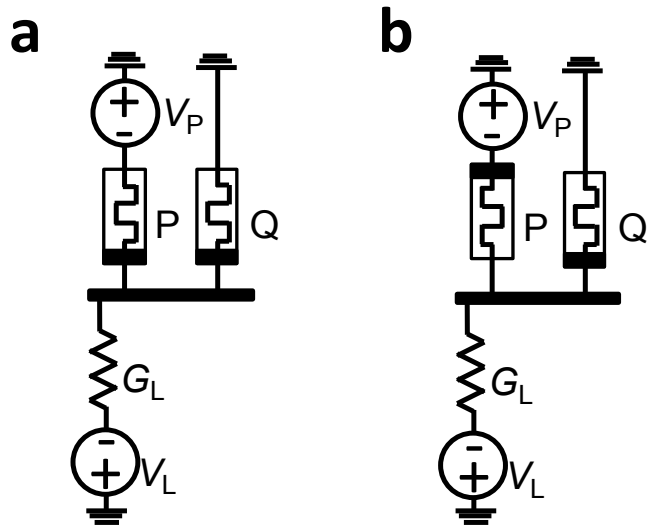

**Figure S6.** (a) Parallel and (b) anti-parallel polarity configuration for memristor-based IMP logic.

Equations 1, 2, 4, and 5 define 12 inequalities in total. To eliminate redundant inequalities, let us first note that $V_L \geq 0$ does not have valid solutions, while $V_P \geq 0$ always results in sub-optimal margins. Assuming $V_P < 0$ and $V_L < 0$ and that memristors P and Q are characterized by the same parameters $V_{set}^{min}$, $V_{set}^{max}$, $V_{reset}^{min}$, $V_{reset}^{max}$, $G_{ON}$, $G_{OFF}$ (a more general case is discussed later) only three conditions must be considered, namely:

- voltage drop on device Q, when Q and P are in the OFF states, is larger than $V_{set}^{max}$,
- voltage drop on device Q, when Q and P are in the ON and OFF states, respectively, is smaller than $V_{set}^{min}$, and
- voltage drop on device P, when Q and P are in the OFF states, is smaller than $V_{set}^{min}$.

Therefore, the largest set margins and the corresponding optimal parameters can be found by solving the following equations:

$$\frac{-V_P G_{OFF} - V_L G_L}{2G_{OFF} + G_L} = V_{set}^* + \Delta_{ideal} \qquad (6)$$

$$\frac{-V_P G_{ON} - V_L G_L}{G_{OFF} + G_{ON} + G_L} = V_{set}^* - \Delta_{ideal} \qquad (7)$$

$$\frac{V_P(G_{ON} + G_L) - V_L G_L}{2G_{OFF} + G_L} = V_{set}^* - \Delta_{ideal} \qquad (8)$$

where

$$V_{set}^* = (V_{set}^{max} + V_{set}^{min})/2 \qquad (9)$$

Here, $\Delta_{ideal}$ is a set margin for the binary zero-variations (i.e. ideal for the considered application) memristors for which $V_{set}^* = V_{set}^{max} = V_{set}^{min}$. Accounting for variations in set switching threshold and analog switching, a more relevant for our case margin is

$$\Delta = \Delta_{ideal} - (V_{set}^{max} - V_{set}^{min})/2 \qquad (10)$$

From Eqs. (7-9) $V_P$, $V_L$ and $\Delta_{ideal}$ are

$$\Delta_{ideal} = V_{set}^* \frac{G_{ON} - G_{OFF}}{2G_L + 3G_{ON} + G_{OFF}} \qquad (11)$$

$$V_P = -2\Delta_{ideal} \qquad (12)$$

$$V_L = -2V_{set}^* \frac{G_L{}^2 + 2G_L(G_{ON} + G_{OFF}) + G_{OFF}(3G_{ON} + G_{OFF})}{G_L(2G_L + 3G_{ON} + G_{OFF})} \tag{13}$$

According to Eq. 10 $\Delta_{ideal}$ is monotonically decreasing with $G_L$ (Fig. 1e) and the maximum margins are achieved for $G_L = 0$, i.e. a circuit on Fig. 1d for which

$$\Delta_{ideal} = V_{set}^* \frac{G_{ON}/G_{OFF} - 1}{3G_{ON}/G_{OFF} + 1} \tag{14}$$

$$I_L \equiv V_L G_L = -2V_{set}^* G_{OFF} \tag{15}$$

For devices with large ON-to-OFF conductance ratio, Eq. 13 can be approximated with very simple formula

$$\Delta_{ideal} \approx V_{set}^*/3. \tag{16}$$

It is instructive to compare IMP logic margins with those of passive crossbar memories. For example, let us consider the most optimal $V/3$-biasing scheme,[1] and assume that voltages $V$ and 0 are applied on the lines leading to the selected device, and $V/3$, and $2V/3$ on the corresponding lines leading to the remaining devices. Assuming that voltage across the selected device is $V = V_{set}^* + \Delta_{memory}$, while it is $V/3 = V_{set}^* - \Delta_{memory}$ across all other devices, it is straightforward to show that the margins for crossbar memory are

$$\Delta_{memory} = V_{set}^*/2. \tag{17}$$

Thus voltage margins for memory circuits are more relaxed as compared to those of IMP logic. In principle, a somewhat larger IMP logic set margins can be obtained by not enforcing full switching, e.g. by defining $V_{set}^{max}$ as the largest set threshold voltage due to cycle-to-cycle variations. However, in this case, the ON-to-OFF ratio will get reduced with every IMP logic operation, which is not desirable.

The analysis above is for a specific IMP logic based on memristors with identical linear static $I$-$V$ characteristics. It is straightforward to extend it to a more general case by using specific to memristors Q and P parameters in Eqs. (S6-S8), such as different set and reset threshold voltages for the top and bottom devices, which is the case relevant to the implemented circuit. For example, a more general set of equations for parallel configuration shown on Fig. S6a, which is more convenient to solve for $\Delta$ directly, is

$$\frac{-V_{\mathrm{P}}G_{\mathrm{OFF}}-V_{\mathrm{L}}G_{\mathrm{L}}}{2G_{\mathrm{OFF}}+G_{\mathrm{L}}} = V_{\mathrm{Q\,set}}^{\max} + \Delta, \quad \frac{-V_{\mathrm{P}}G_{\mathrm{ON}}-V_{\mathrm{L}}G_{\mathrm{L}}}{G_{\mathrm{OFF}}+G_{\mathrm{ON}}+G_{\mathrm{L}}} = V_{\mathrm{Q\,set}}^{\min} - \Delta, \quad \frac{V_{\mathrm{P}}(G_{\mathrm{ON}}+G_{\mathrm{L}})-V_{\mathrm{L}}G_{\mathrm{L}}}{2G_{\mathrm{OFF}}+G_{\mathrm{L}}} = V_{\mathrm{P\,set}}^{\min} - \Delta \quad (18)$$

from which the actual margin for $G_{\mathrm{L}}= 0$  is

$$\Delta = \frac{(G_{\mathrm{ON}}+G_{\mathrm{OFF}})\left(V_{\mathrm{Q\,set}}^{\min}-V_{\mathrm{Q\,set}}^{\max}\right) + (G_{\mathrm{ON}}-G_{\mathrm{OFF}})V_{\mathrm{P\,set}}^{\min}}{3G_{\mathrm{ON}}+G_{\mathrm{OFF}}} \quad (19)$$

For anti-parallel configuration shown on Fig. S6b, the set of equation is

$$\frac{-V_{\mathrm{P}}G_{\mathrm{OFF}}-V_{\mathrm{L}}G_{\mathrm{L}}}{2G_{\mathrm{OFF}}+G_{\mathrm{L}}} = V_{\mathrm{Q\,set}}^{\max} + \Delta, \quad \frac{-V_{\mathrm{P}}G_{\mathrm{ON}}-V_{\mathrm{L}}G_{\mathrm{L}}}{G_{\mathrm{OFF}}+G_{\mathrm{ON}}+G_{\mathrm{L}}} = V_{\mathrm{Q\,set}}^{\min} - \Delta, \quad \frac{V_{\mathrm{P}}(G_{\mathrm{ON}}+G_{\mathrm{L}})-V_{\mathrm{L}}G_{\mathrm{L}}}{2G_{\mathrm{OFF}}+G_{\mathrm{L}}} = -(V_{\mathrm{P\,set}}^{\min} - \Delta) \quad (20)$$

and the actual margin for $G_{\mathrm{L}}= 0$ is

$$\Delta^{\mathrm{anti}} = \frac{(G_{\mathrm{ON}}+G_{\mathrm{OFF}})(V_{\mathrm{Q\,set}}^{\min}-V_{\mathrm{Q\,set}}^{\max})- (G_{\mathrm{ON}}-G_{\mathrm{OFF}})V_{\mathrm{P\,reset}}^{\min}}{3G_{\mathrm{ON}}+G_{\mathrm{OFF}}} \quad (21)$$

Because $-V_{\mathrm{reset}}^{\min} > V_{\mathrm{set}}^{\min}$ typically holds for the considered devices (Fig. S5), from Eqs. 19 and 21 margins for the parallel case are smaller, which is why this case is considered more in detail. Margins and optimal parameters for the remaining parallel (Fig. 4a) and antiparallel configurations (Fig. 4d) that were experimentally demonstrated, are similar to those described above with the only difference is that the signs for $V_{\mathrm{P}}$ and $I_{\mathrm{L}}$ are negative.

The analytical approach can be also utilized for IMP logic based on memristors with more realistic nonlinear static $I$-$V$ by using $G_{\mathrm{ON}}$ and $G_{\mathrm{OFF}}$ measured at large (close to switching threshold) voltages. A more accurate approach, however, is to solve inequalities Eqs. (S1-S5) numerically. By fitting experimental $I$-$V$ curves (Fig. S4b) and using Mathematica's Newton-Raphson-based solver, we have obtained more accurate optimal values for $V_{\mathrm{P}}$ and $V_{\mathrm{L}}$, which were used in experimental work.  The margins calculated from a numerical simulations for a specific IMP logic are also shown on Fig. 1e and are in fairly good agreement with simple analytical model for a system with an ON-to-OFF conductance ratio of ~10.

## 4. Material implication logic experiment

For IMP and NAND experiments presented in Figs. 4 and 5, the memristors were set to the initial states using the state tuning algorithm.[2] In particular, a train of 1-ms pulses with increasing amplitude, starting from 0.5 V to maximum of 1.9 V with 0.1 V steps for reset pulses, and from 50 µA to 900 µA with 50 µA step for set pulses. Initial state ON and OFF conductances measured at 0.1 V were always close to 115 μS, 115 μS, 125 μS, 120 μS and 10 μS, 10 μS, 5 μS, 8 μS for B1, B2, T1, and T2 devices, correspondingly. The optimal $V_P$ and $I_L$ were determined from numerical simulations with an additional constraint of using the same circuit parameters when the IMP logic output is in the bottom or top memristors. Such additional constraint is representative of a more general case when parameters of the biasing circuitry are not chosen based on switching characteristics of individual memristors.

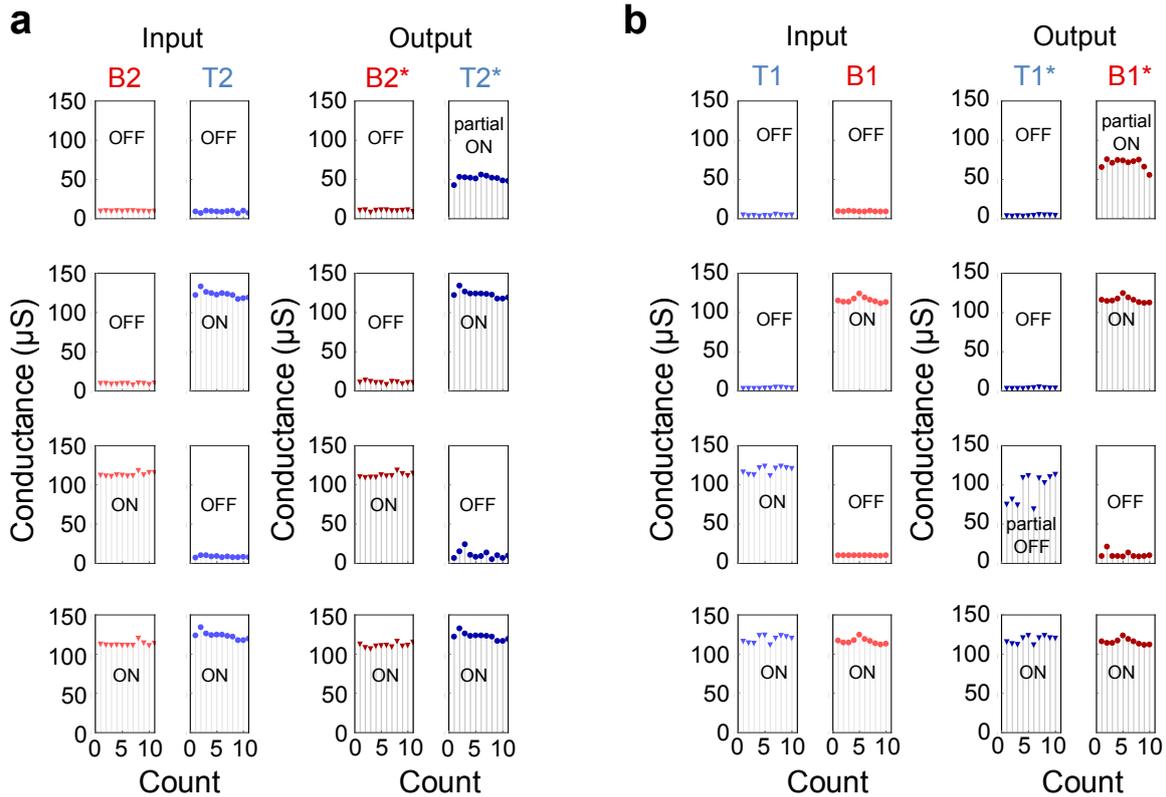

**Figure S7.** Detailed information for 10 representative cycles for (a) T2*← B2 IMP T2 and (b) T2*← B2 IMP T2.